\documentclass[aps,prb,twocolumn,showpacs,superscriptaddress]{revtex4-1}
\usepackage[dvips,pdftex]{graphicx}
\usepackage{hyperref}
\usepackage{amsmath}
\usepackage{ulem}
\usepackage{color}



\begin{document}
\title{
Quantum chaotic subdiffusion in random potentials}
\author{M.V.~Ivanchenko}
\email[]{ivanchenko@rf.unn.ru}
\affiliation{Department for Bioinformatics, Lobachevsky State University of Nizhny Novgorod, Russia}

\author{T.V.~Laptyeva}
\affiliation{Theory of Control Department, Lobachevsky State University of Nizhny Novgorod, Russia}

\author{S.~Flach}
\affiliation{New Zealand Institute for Advanced Study, Massey University, Auckland, New Zealand}

\date{\today}

\begin{abstract}
Two interacting particles (TIP) in a disordered chain propagate beyond the single particle localization length $\xi_1$ up to a scale $\xi_2 > \xi_1$. An initially strongly localized TIP state expands almost ballistically up to $\xi_1$. The expansion of the TIP wave function beyond the distance $\xi_1 \gg 1$ is governed by highly connected Fock states in the space of noninteracting eigenfunctions. The resulting dynamics is subdiffusive, and the second moment grows as $m_2 \sim t^{1/2}$, precisely as in the strong chaos regime
for corresponding nonlinear wave equations. This surprising outcome stems from the huge Fock connectivity and resulting quantum chaos. The TIP expansion finally slows down towards a complete halt -- in contrast to the nonlinear case. 
\end{abstract}


\maketitle

Anderson localization (AL), the absence of diffusion in linear lattice wave equations due to disorder \cite{Anderson}, is now broadly seen as a fundamental physical phenomenon manifested by light, sound, and matter waves \cite{Experiments,Evers}. Rigorous results state that in one dimension all single-particle (SP) states become exponentially localized at arbitrary weak disorder \cite{Gert,Abrahams,Kramers}. 
Going beyond the assumption of non-interacting particles or linear waves has proved to be extremely complex, and the current answers on the interplay between disorder and interactions remain controversial and debated. 

For quantum many body systems predictions range from no major effect of interactions \cite{Anderson80} to the emergence of a finite-temperature AL transition already in dimension one \cite{Basko}. Advance has been achieved within mean field approximations, which lead to nonlinear wave equations like the Gross-Pitaevsky equation \cite{Dalfovo}. Recent analytical and numerical studies demonstrate that nonlinearity breaks AL and leads to subdiffusive wave packet propagation, caused by nonintegrability, deterministic chaos, phase decoherence and a consequent loss of wave localization \cite{molina,Pikovsky_Shepelyansky,Flach09,laptyeva10,bodyfelt11,Pikovsky,larcher12}. A positive measure of initially localized excitations gets delocalized for an arbitrarily small nonlinearity, tending to one above some threshold which may depend on the initial state \cite{Magnus}.  

Few interacting quantum particles may bridge the two extremes from above. In particular the case of two interacting particles (TIP) appears to be an interesting testing ground for any of the above statements. There is not much doubt that the TIP case also yields a finite localization length $\xi_2$, similar to but potentially much larger than the single particle localization length $\xi_1$. Most of the studies only debate whether and how the TIP localization length $\xi_2$ scales with $\xi_1$
\cite{Dorohov,Shep94,Imry,Roemer,Frahm,Krimer11}. Paradoxically, almost nothing is known on the interaction induced wave packet dynamics beyond $\xi_1$. Numerical experiments explored only the case of the strong disorder, limiting to small connectivity of TIP states in the relevant Hilbert space of the problem, to report, unsurprisingly, quite a trivial ballistic expansion up to $\xi_1$ followed by a quick saturation \cite{Arias}.

Here we report the first study of TIP wave packet dynamics much beyond the single particle localization length and discover the new phenomenon of quantum chaotic subdiffusion. This is exhibited in the weak disorder regime, characterized by the large connectivity of Fock states, when the packet size exceeds $\xi_1$ but does not yet reach its asymptote $\xi_2$. We find that in this regime the system demonstrates the two main signatures of quantum chaos: dynamical excitation of a wealth of Fock states and strongly non-Poissonian level spacing distribution. Based on this, we argue that the TIP subdiffusion in Fock and in real space can be described as subsequent excitation of unpopulated Fock states by a multifrequency and quasistochastic driving from a large number of already populated ones. Numerics and analysis estimate the subdiffusion exponent as $1/2$ in a remarkable correspondence to the strong chaos subdiffusion in the classical nonlinear wave equation. Asymptotic dynamics differs though, in the crossover to a weak chaos subdiffusion of the nonlinear case, and slowing down towards a complete arrest of the TIP excitation. 

We study the TIP dynamics in the framework of the Hubbard model with the Hamiltonian
\begin{equation}
\label{eq1}
{\cal \hat H}= \sum\limits_j\left[\hat b_{j+1}^+\hat b_j+\hat
b_{j}^+\hat b_{j+1}+\epsilon_j\hat b_{j}^+\hat b_j+\frac{U}{2}\hat
b_{j}^+\hat b_j^+\hat b_{j}\hat b_j\right]
\end{equation}
where $\hat b_{j}^+$ and $\hat b_{j}$ are creation and
annihilation operators of indistinguishable bosons at lattice site
$j$, and $U$ measures the on-site interaction strength between the particles.
The on-site energies are random uncorrelated numbers with a uniform probability density function on the interval $\epsilon_j\in[-W/2,W/2]$ as in the original Anderson problem.

Using the vacuum state $|0\rangle $ and the basis $|j,k\rangle\equiv\hat b_j^+b_k^+|0\rangle$ we 
write the TIP wave function as $\Psi=\sum\limits_{j,k}\varphi_{j,k}|j,k\rangle$. 
Note that indistinguishability implies $j \geq k$. However this can be eased by considering 
arbitrary pairs of $j,k$ with the constraint that $\varphi_{j,k}=\varphi_{k,j}$.
Inserting into the 
Schroedinger equation $i\dot{\Psi} = {\cal \hat H} \Psi$ we  
obtain an effective single particle problem on a two-dimensional lattice with correlated disorder and
a defect line along the diagonal due to the interaction $U$ (here $\hbar=1$):
\begin{equation}
\label{eq2}
i\dot{\varphi}_{j,k}=\epsilon_{j,k}\varphi_{j,k}+\sum\limits_\pm(\varphi_{j,k\pm1}+\varphi_{j\pm1,k}),
\end{equation}
where $\epsilon_{j,k}=\epsilon_j+\epsilon_k+U\delta_{j,k}$ and $\delta_{j,k}$ is the Kronecker symbol.
In the absence of interactions $U=0$ solutions to (\ref{eq2}) break into a product of SP solutions that follow
\begin{equation}
\label{eq3}
i\dot{\varphi}_{m}=\epsilon_{m}\varphi_{m}+\varphi_{m+1}+\varphi_{m-1}.
\end{equation} 
All eigenstates $\{A^{(r)}_m\}_{r=1,\ldots N}$ of (\ref{eq3}) are exponentially localized with the maximal localization length $\xi\approx\frac{96}{W^2}$ for $W < 4$ \cite{Kramers}. Their corresponding eigenvalues are
$\omega_{r}$.

Let us rewrite the dynamical equations (\ref{eq2}) in the reciprocal Fock space of SP eigenstates $\phi^{(r_1,r_2)}=\sum_{j,k} A^{(r_1)}_j A^{(r_2)}_k \varphi_{j,k}$:
\begin{equation}
\label{eq4}
i\dot{\phi}^{(r_1,r_2)}=\omega_{r_1,r_2}\phi^{(r_1,r_2)}+U\sum\limits_{s_1,s_2} I_{r_1,r_2,s_1,s_2} \phi^{(s_1,s_2)},
\end{equation}
where $\omega_{r_1,r_2}=\omega_{r_1}+\omega_{r_2}$ is the sum of SP eigenvalues $\omega_{r_1}$ and $\omega_{r_2}$, and $I_{r_1,r_2,s_1,s_2}=\sum_j A^{(r_1)}_j A^{(r_2)}_j A^{(s_1)}_j A^{(s_2)}_j$ is the
overlap integral.

The interaction $U$ transfers excitation amplitudes between two-particle Fock states. Suppose that the particles initially occupy a Fock state $(s_1,s_2)$. 
For $U=0$ the solution to (\ref{eq4}) reads $\phi^{(s_1,s_2)}=\phi_0 e^{-i \omega_{s_1,s_2}t}$. 
For nonzero $U$, in the first order the other states follow
\begin{equation}
\label{eq5}
\phi^{(r_1,r_2)}\sim \frac{U I_{r_1,r_2,s_1,s_2}}{\omega_{r_1,r_2}-\omega_{s_1,s_2}} \phi_0 e^{i \omega_{s_1,s_2}t}.
\end{equation}  
If the resonance condition 
\begin{equation}
\label{eq6}
\left|\frac{U I_{r_1,r_2,s_1,s_2}}{\omega_{r_1,r_2}-\omega_{s_1,s_2}}\right|\sim1
\end{equation} 
is fulfilled, the states $(r_1,r_2)$ and $(s_1,s_2)$ become hybridized. Note that they will partially occupy the same volume since the overlap integrals of distant Anderson modes are exponentially small. Similar arguments can be mounted in higher orders of perturbation theory. 
The inverse characteristic time for the hybridization to take place is of the order $U I_{r_1,r_2,s_1,s_2}$.

An increase of the interaction strength $U$ appears to increase the number of potential hybridizations and facilitates the dynamical process. However this is only true as long as $U$ is less or equal than the
single particle kinetic energy which is of order one in Eq.(\ref{eq1}). For $U \gg 1$ the TIP spectrum splits into a noninteracting spinless fermion continuum and a band of double occupied site states, separated from the continuum by energy $\sim U$. The spinless fermions will yield a localization length very close to $\xi_1$ since they do not interact
(apart from avoiding double occupancy). The separated states will have a much smaller localization length
of the order of $\xi_1/U^2$. Therefore the optimal value for the interaction is $U \sim 1$.

At variance to the many body problem, or the nonlinear wave case, where the number of particles or the density serve as an additional control parameter, we are then left in the TIP case with only one control parameter - the disorder strength $W$. As there are about $\xi_1^2$ Fock states residing in a volume of size $\xi_1$,
their connectivity, if at all, will be large for weak disorder.  It is the weak disorder case $\xi_1 \gg 1$ which we will therefore
explore. 

We study the expansion of an initial TIP wavepacket whose
size is substantially smaller than $\xi_1$. A single particle wave packet will expand ballistically up to 
$\xi_1$ which serves both as a localization length and as a mean free path in one dimension. 
We expect therefore that also a TIP packert will expand ballistically (or at least faster than diffusive)
up to the length scale $\xi_1$. Beyond that length any further expansion will be due to the interaction
and the nontrivial dynamics in Fock space.
We perform extra-large-scale computational studies of TIP dynamics (\ref{eq2}) for $W<2$, where a substantial enhancement of $\xi_2$ is expected \cite{Krimer11}. We analyze the wave packet expansion and develop a qualitative theoretical explanation of our observations.

\begin{figure}
\begin{center}
\includegraphics[width=\columnwidth,keepaspectratio,clip]{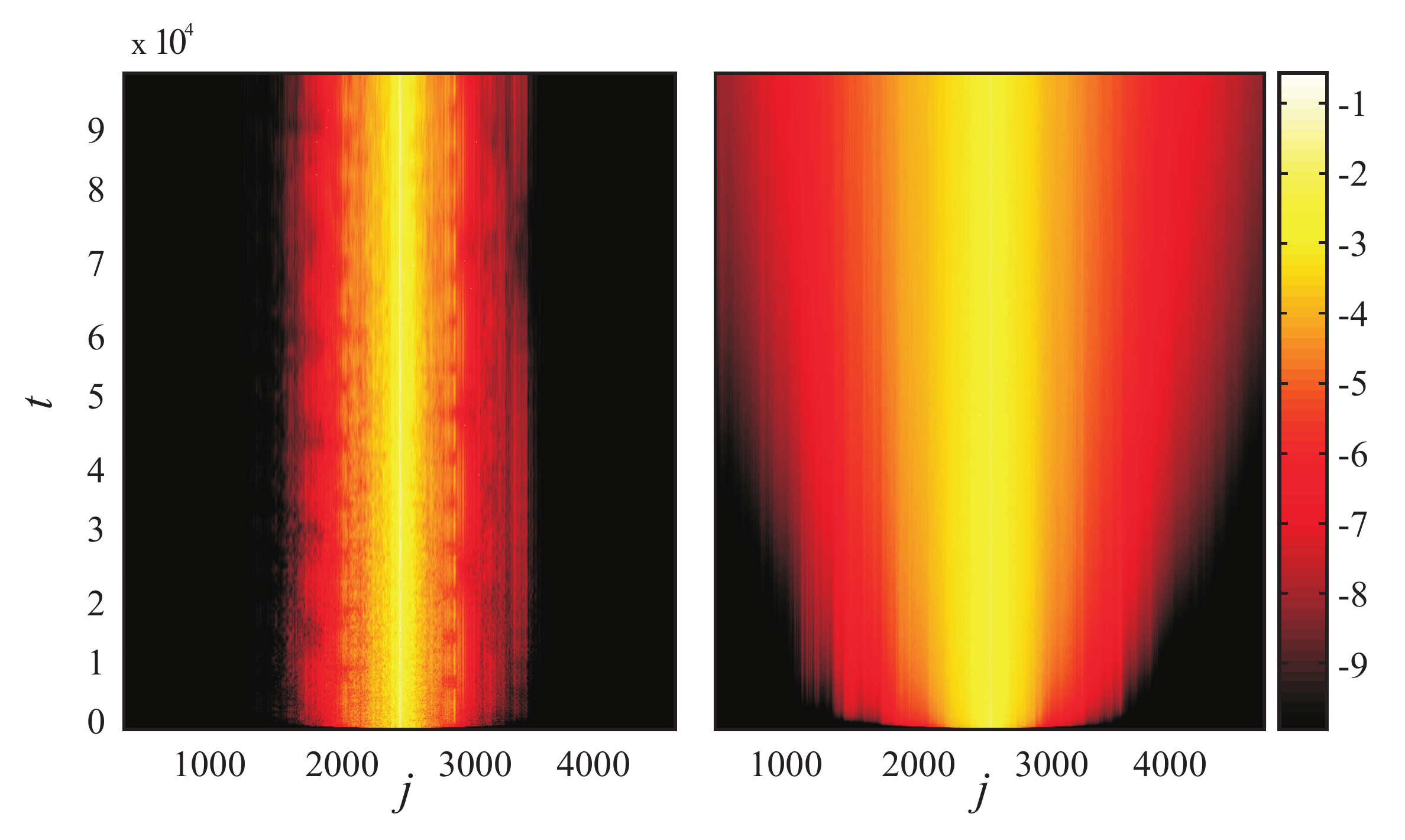}
\caption{Time-space dependence of the PDF $\log_{10} z_j$ of an initial state for $W = 1.0$, $U = 0.0$ (left panel) and $U = 2.0$ (right panel). The color code shows $\log_{10} z_j$ for a particular disorder realization.}
\label{fig:PDFplots}
\end{center}
\end{figure}

\begin{figure}
\begin{center}
\includegraphics[width=\columnwidth,keepaspectratio,clip]{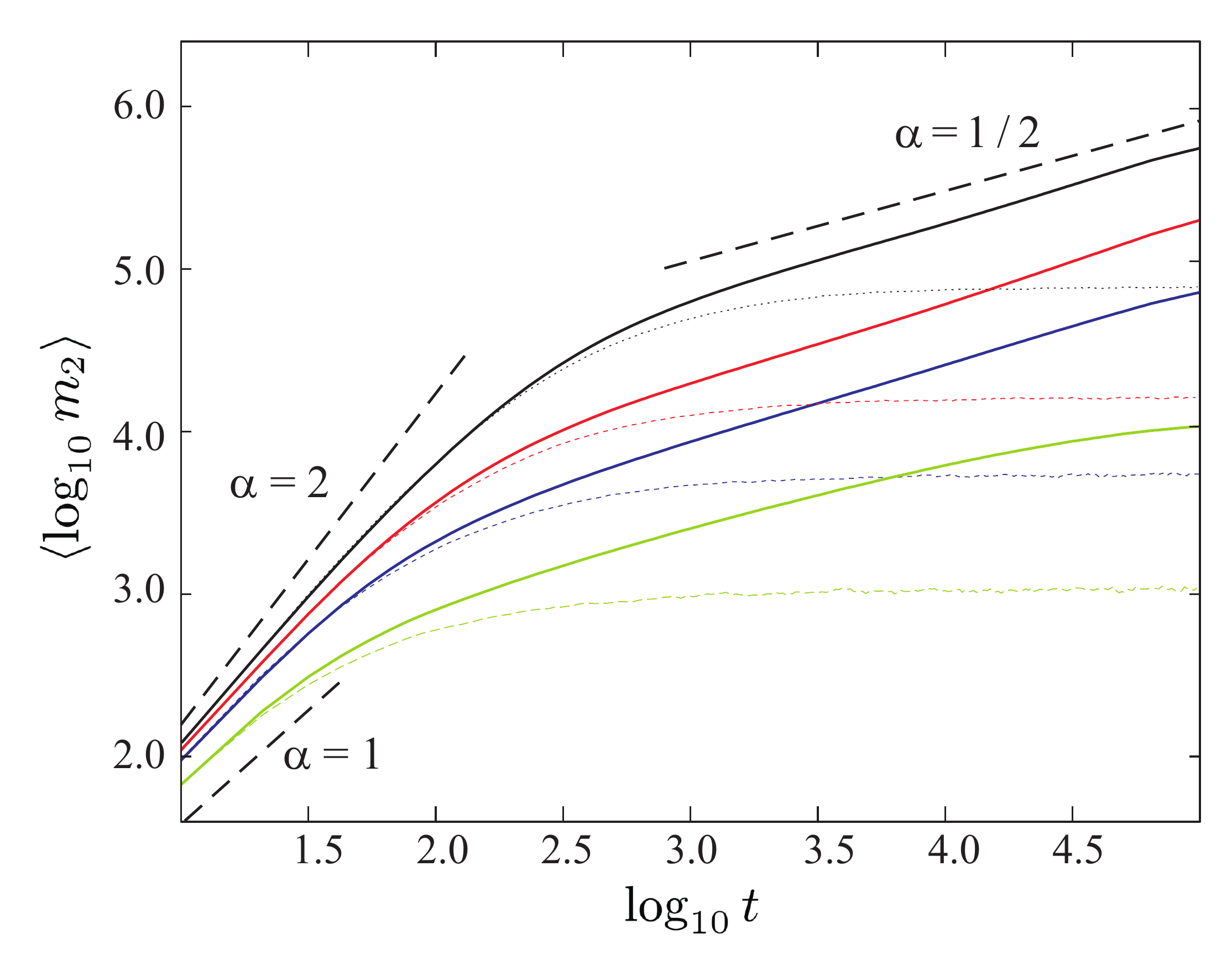}
\caption{(Color online) $\langle \log_{10} m_2 \rangle$ versus $\log_{10} t$ for $U = 2.0$ (solid lines) and $U = 0$ (dotted lines). The four curves (from top to bottom) correspond to $W = 0.5$ (black), $W = 0.75$ (red), $W = 1.0$ (blue), and $W = 1.5$ (green). Black solid lines are to guide the eye to the values $\alpha = 1$, $\alpha = 2$, and $\alpha = 1/2$ (see marks in the figure).
}
\label{fig:ssite}
\end{center}
\end{figure}

\begin{figure}
\begin{center}
\includegraphics[width=\columnwidth,keepaspectratio,clip]{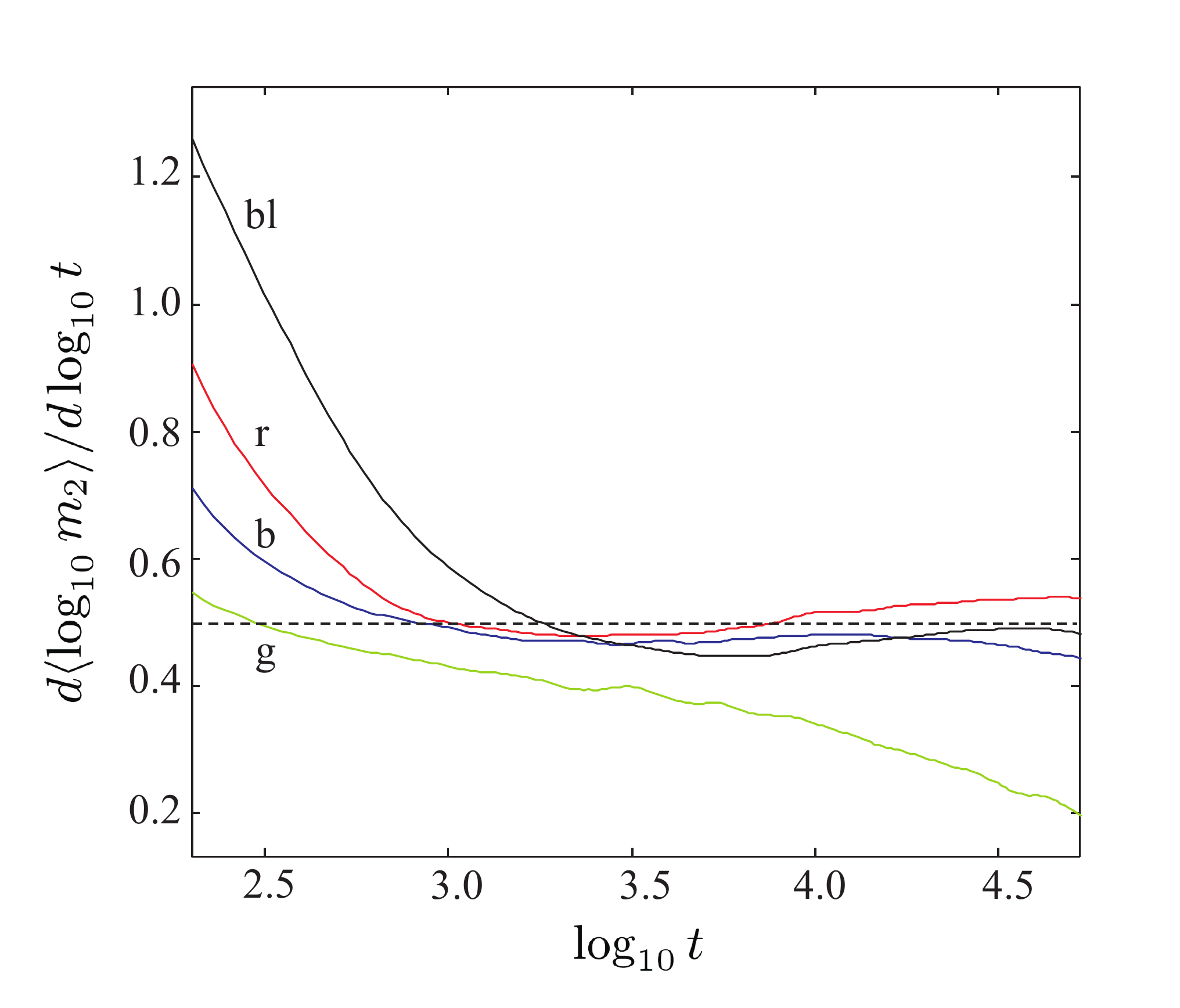}
\caption{(Color online) Local derivative $\alpha={\rm d}\langle\log_{10}m_2\rangle/{\rm d}\log_{10}t$ for the data from Fig.\ref{fig:ssite}: the four curves correspond to $W = 0.5$ (bl - black), $W = 0.75$ (r - red), $W = 1.0$ (b - blue), and $W = 1.5$ (g - green). The dashed horizontal line corresponds to value $\alpha = 1/2$.}
\label{fig:msite}
\end{center}
\end{figure}

Numerical integration of (\ref{eq2}) is performed on a finite lattice $N\times N$ with the PQ-method \cite{bodyfelt11}. The two particles are initially placed at neighbour sites. The probability distribution function (PDF) of the particle density is given by 
$P_j=\sum\limits_k |\varphi_{j,k}|^2$, and normalized to $z_j=P_j/\sum_k P_k$.
We monitor the wave packet expansion computing its mass center $m_1=\sum_j j z_j$ and the second moment $m_2=\sum_j(j-m_1)^2 z_j$. For each choice of parameters $W$ and $U$ we average the numerical data over 100 different disorder realizations and denote this by $\left\langle \cdots \right\rangle$. 
In most of experiments we set the system size $N=5000$, making also test simulations with $N=8000$ to make sure that boundary effects do not matter. 

The evolution of the two-particle PDF in the weak disorder regime is presented in Fig.~\ref{fig:PDFplots} for $W = 1.0$. In the non-interacting case we observe a rapid expansion of a wave packet over the SP localization volume and a halt afterwards ($U = 0$, left panel). At variance, interactions promote the wave packet diffusion beyond the SP localization volume and we do not observe visual signs of its halt up to $t=10^5$, about two orders of magnitude beyond the SP expansion time  ($U = 2.0$, right panel). 

In Fig.~\ref{fig:ssite} we plot the evolution of the corresponding second moment $m_2$.
In a range of disorder strength $W = 0.5 \ldots 1.5$ we observe that after an initial superdiffusive
spread (a mix of ballistic transport and the influence of interaction) the spreading turns into
a potentially long lasting subdiffusive regime.
A comparison to the non-interacting case ensures that the discovered subdiffusion is due to interactions (Fig.~\ref{fig:ssite}). 

In order to quantify our findings, we first smooth $\left\langle \log_{10} m_2\right\rangle$
with a locally weighted regression algorithm \cite{smoothing}, and then apply a central finite-difference to calculate the local derivative  $\alpha = d \langle \log_{10} m_2 \rangle/ d\log_{10} t$ for the data from Fig.\ref{fig:ssite} and plot the result in Fig.\ref{fig:msite}. 
For $W=1.5$ the ballistic/superdiffusive spreading continuously slows down as time progresses.
However for weaker disorder $W=1,0.75,0.5$ the ballistic/superdiffusive regime crosses over into
a subdiffusive one with $\alpha \approx 0.5$, and that regime lasts at least for 1.5 decades. 
This is precisely what is known as the strong chaos regime for the chaotic spreading of nonlinear wave packets
in disordered potentials \cite{laptyeva10}.

\begin{figure}
\begin{center}
\includegraphics[width=\columnwidth,keepaspectratio,clip]{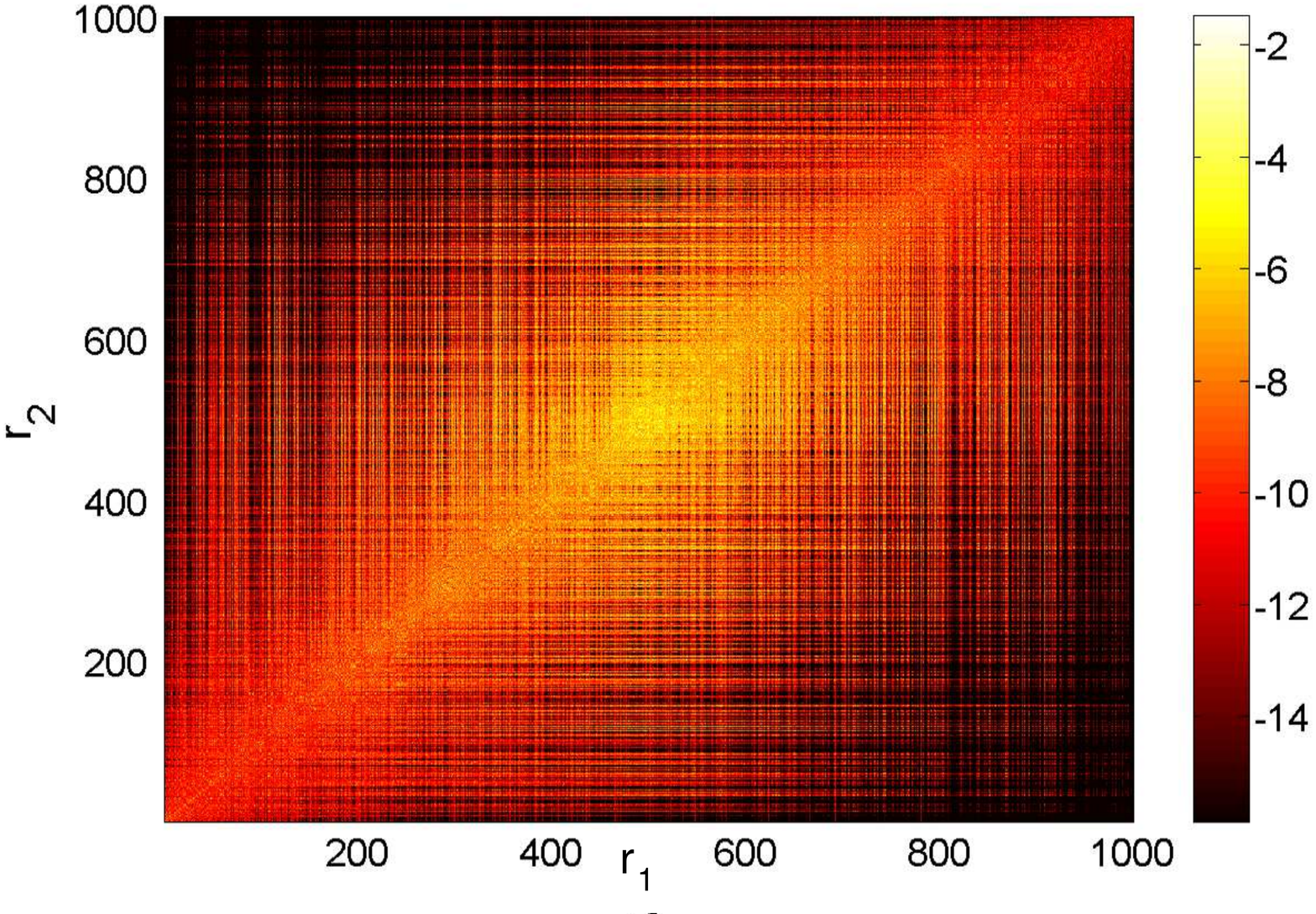}
\caption{Distribution of norm in Fock space at $t=500$ after an evolution from a single initial Fock state for $W = 1.0$, $U = 2.0$. The color code shows $\log_{10}{\phi^{(r_1,r_2)}}$.}
\label{fig:Fig4}
\end{center}
\end{figure}

\begin{figure}
\begin{center}
\includegraphics[width=\columnwidth,keepaspectratio,clip]{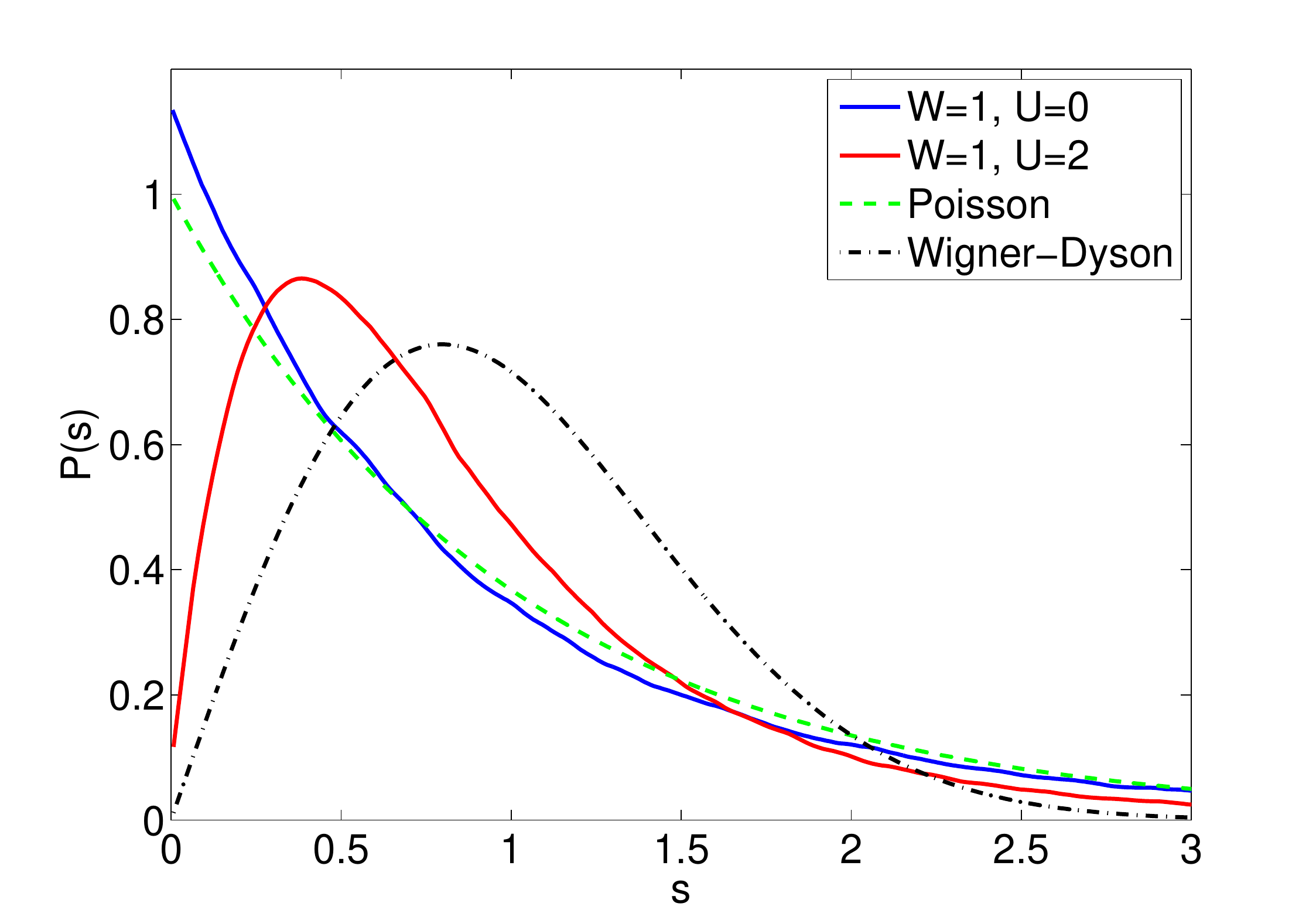}
\caption{Normalized TIP level spacing distribution ($\left\langle P(s)\right\rangle=1$) for $W=1$ and $N=100$: close to Poisson in absence of interactions ($U = 0$, blue solid line), and distinctly non-Poissonian ($U=2$, red solid line), with the sub-linear level repulsion fit $P(s)\propto s^\beta, \ \beta\approx0.6, \ s\ll 1$. Testbed Poisson $P(s)=\exp{(-s)}$ (green dashed line) and Wigner-Dyson $\pi s/2\exp{(-\pi s^2 /4)}$ (black dash-dotted line) distributions are also shown.}
\label{fig:Fig5}
\end{center}
\end{figure}

We conjecture and demonstrate that this similarity has a profound physical origin in quantum chaos, its two main signatures displayed \cite{quantum_chaos}. First, the weak disorder regime leads to high connectivity of the Fock states. In Fig.\ref{fig:Fig4} we plot the norm distribution in the Fock space after an initial excitation of a single state. Clearly a huge number of the other states become populated. Second, the normalized TIP level spacing distribution $P(s)$ becomes strongly non-Poissonian for the set of parameters that exhibits subdiffusion (cf. Fig.\ref{fig:ssite}, \ref{fig:msite}), as opposed to an almost Poisson law for zero interactions (cf. Fig.\ref{fig:Fig5} illustrating the case $W=1, \ U=0$ and $U=2$, measured on the block about $\xi_1\times\xi_1$). Here level repulsion reveals the parameter-dependent sub-linear fit $P(s)\propto s^\beta, \ \beta<1, \ s\ll 1$. The resulting quantum chaotic oscillations determine the proximity to nonlinear chaotic dynamics on the timescale of their inverse average frequency spacing.


The corresponding diffusion rate $D$ is proportional to the coupling $\Gamma_{\mathbf{r},\mathbf{s}}$ between the initial and final states $\mathbf{r}\equiv(r_1,r_2), \ \mathbf{s}\equiv(s_1,s_2)$. Perturbative calculations give \cite{Lucioni2013}:
\begin{equation}
\label{eq7}
\Gamma_{\mathbf{r},\mathbf{s}}=2\pi\frac{|\left\langle \mathbf{r} | \mathcal{H}_{int} | \mathbf{s} \right\rangle|^2}{|\omega_\mathbf{r}-\omega_\mathbf{s}|}, \ \mathcal{H}_{int}=\sum\limits_j \frac{U}{2} \hat
b_{j}^+\hat b_j^+\hat b_{j}\hat b_j,   
\end{equation}
and yield $D\sim U^2 \mathbf{n}^2$, where $\mathbf{n}$ is the local norm density in Fock space. Since we follow
the wave packet in a one-dimensional system, it follows that $m_2\sim 1/\mathbf{n^2}$. Substituting the above into $m_2=D~t$ we finally arrive at
\begin{equation}
\label{eq8}
m_2 \propto t^{1/2},  
\end{equation}
which corresponds well to our numerical findings.

In conclusion, we discovered the TIP subdiffusion in the weak disorder regime of Anderson localization and found a remarkable correspondence of the power law exponent $\alpha=1/2$ to the one observed for classical strongly chaotic nonlinear waves \cite{laptyeva10}. We demonstrated the pronounced signatures of quantum chaos in this regime, and proposed a mechanism conjecturing the origin of subdiffusion from the quantum chaos of strongly interacting two-particle Fock states. The obtained results call for further research to provide a rigorous description of the mechanisms behind TIP subdiffusion, to explore the relation to the random walk hopping theory of subdiffusion in disordered semiconductors \cite{Scher1975}, and put a thrilling question on how the asymptotic $\alpha=1/3$ power law subdiffusion \cite{Flach09} converging to a self-similar solution for nonlinear waves \cite{nde} is recovered in a quantum system with $N > 2$ particles. It would be also extremely interesting to address these phenomena in experiments using interacting pairs of ultracold Rb atoms in optical lattices \cite{lucioni,immbloch}, employing the recent advances in single atom control \cite{bloch2}.    

MVI and TVL acknowledge financial support of RF President grant MK-4028.2012.2 and RFBR 12-02-31403. Dynasty Foundation is also acknowledged. A significant part of numerical experiments has been carried out at the HPC of Lobachevsky State University of Nizhny Novgorod. 
 

\end{document}